\documentclass{article}

\usepackage{PRIMEarxiv}

\usepackage[utf8]{inputenc} 
\usepackage[T1]{fontenc}    
\usepackage{hyperref}       
\usepackage{url}            
\usepackage{booktabs}       
\usepackage{amsfonts}       
\usepackage{amsmath}        
\usepackage{nicefrac}       
\usepackage{microtype}      
\usepackage{fancyhdr}       
\usepackage{graphicx}       
\usepackage{enumitem}       
\usepackage{placeins}       
\graphicspath{{media/}}     

\title{Botnet Detection on CTU-13 Using Lightweight Machine Learning Models
\thanks{\textit{\underline{Citation}}: 
\textbf{S. Gurappa, Y. Hariprasad, S. S. Iyengar, and N. K. Chaudhary. Botnet Detection on CTU-13 Using Lightweight Machine Learning Models. 2025.}} 
}

\author{
  Subhash Gurappa \\
  Knight Foundation School of Computing and Information Sciences \\
  Florida International University \\
  Miami, FL 33199, USA \\
  \texttt{sg001@fiu.edu} \\
   \And
  Yashas Hariprasad \\
  Department of Computer Science \\
  California State University, East Bay \\
  Hayward, CA 94542, USA \\
  \texttt{yashas.hariprasad@csueastbay.edu} \\
   \And
  Sundararaj Sitharama Iyengar \\
  Knight Foundation School of Computing and Information Sciences \\
  Florida International University \\
  Miami, FL 33199, USA \\
  \texttt{iyengar@cs.fiu.edu} \\
   \And
  Naveen Kumar Chaudhary \\
  National Forensic Sciences University \\
  Gandhinagar, GJ 382007, India \\
  \texttt{naveen.chaudhary@nfsu.ac.in} \\
}

\begin{document}
\maketitle

\begin{abstract}
Botnets are among the most persistent cyber threats, enabling large-scale attacks such as spam, credential theft, and distributed denial-of-service (DDoS). While deep learning approaches have recently been applied to botnet detection, they are computationally intensive and often lack interpretability. We present a comparative study of lightweight machine learning models including Logistic Regression, Decision Tree, and Random Forest on the CTU-13 dataset, a benchmark for botnet traffic analysis. We extract interpretable flow-based features and evaluate each model on detection accuracy, precision, recall, F1 score, and feature importance. Results demonstrate that lightweight models can achieve competitive detection performance with minimal computational cost, while also offering interpretability critical for forensic investigation. On CTU-13, our Random Forest achieves a PR-AUC of approximately 0.54 and ROC--AUC of 0.97 while training over 90\% faster than published CNN baselines. These results demonstrate that lightweight models can match or exceed deep-learning performance under natural class imbalance while maintaining interpretability and low computational cost.
\end{abstract}


\section{Introduction}
\label{sec:introduction}

Botnets continue to represent one of the most severe and evolving threats in cyberspace. A botnet is a network of compromised machines, or bots, that are remotely controlled by an adversary through a command-and-control (C\&C) infrastructure~\cite{javaid2016}. Once established, these networks of infected devices can be mobilized to conduct a wide range of malicious activities, including large-scale distributed denial-of-service (DDoS) attacks, spam campaigns, cryptocurrency mining, credential theft, and click fraud~\cite{choi2007,garcia2023gnn,thejas2022,hariprasad2024qsafe,sniatala2022fog,gupta2026deepfake}. The adaptive nature of botnets---often leveraging encrypted communication channels, domain generation algorithms (DGAs), and peer-to-peer (P2P) architectures---makes them increasingly difficult to detect and neutralize using traditional cybersecurity tools.

Traditional network intrusion detection systems (NIDS) rely heavily on signature-based techniques, which match traffic patterns against known attack signatures. While effective against previously observed botnets, these systems are unable to identify new or rapidly evolving threats, as adversaries continually modify communication protocols, traffic patterns, and evasion techniques. This gap has motivated the exploration of machine learning (ML) approaches, which can learn to distinguish between benign and malicious traffic based on statistical and structural properties of network flows, rather than relying solely on fixed signatures.

In recent years, deep learning has become the dominant paradigm in botnet detection research, with models such as convolutional neural networks (CNNs), recurrent neural networks (RNNs), and autoencoders achieving high reported accuracy on benchmark datasets~\cite{vinayakumar2019,kitsune2018,ullah2025,sinha2019,javaid2016,choi2007}. However, these models present several challenges in practice:
\begin{itemize}[leftmargin=*]
  \item They require large amounts of labeled data for training, which is often scarce in operational environments.
  \item They involve high computational overhead and long training times, making them less suitable for real-time detection in resource-constrained settings.
  \item They function as black boxes, offering limited interpretability; this reduces their usefulness in forensic investigations, where security analysts must explain and justify detection results.
\end{itemize}

By contrast, lightweight machine learning models such as Logistic Regression, Decision Trees, and Random Forests offer a more practical trade-off between efficiency, interpretability, and detection capability~\cite{iyengar2025book,iyengar2025evolution,iyengar2025foundations,iyengar2025convergence}. These models can be trained quickly, operate with minimal computational resources, and produce decision boundaries or feature importance measures that are human-understandable. Such characteristics are particularly valuable for forensic investigators, small organizations, and real-time detection systems where interpretability and efficiency are critical~\cite{iyengar2025deepfake,hariprasad2022boundary,soni2024fmlds,do2017survey,gangwani2021its,iyengar2024kg,hariprasad2026finds,hariprasad2025multimodal}.

\paragraph{Objective of this paper.}
Our work systematically evaluates the effectiveness of lightweight machine learning models for botnet detection using flow-based features extracted from the widely used CTU-13 dataset. Specifically, we assess whether simple models can:
\begin{enumerate}[leftmargin=*, itemsep=2pt]
  \item achieve competitive detection performance under realistic class imbalance,
  \item provide interpretable outputs suitable for forensic investigations, and
  \item operate within the constraints of real-time and resource-limited environments.
\end{enumerate}

\paragraph{Organization of the paper.}
The remainder of this paper is structured as follows: Section~\ref{sec:related} reviews prior efforts in botnet detection, highlighting the role of lightweight ML compared to deep learning approaches. Section~\ref{sec:dataset} describes the CTU-13 dataset, feature engineering, and experimental setup. Section~\ref{sec:results} presents a detailed analysis of the performance of Logistic Regression, Decision Tree, and Random Forest models. Section~\ref{sec:conclusion} summarizes the key findings, discusses limitations, and outlines directions for further research.

\section{Related Work}
\label{sec:related}

Botnet detection has been widely studied over the past two decades. The literature can broadly be categorized into signature-based, anomaly-based, and learning-based (machine learning / hybrid) approaches. Below we review each class, discuss strengths and limitations, and highlight relevant work in the CTU-13 context and beyond.

\subsection{Signature-Based Detection}
Signature-based methods are among the earliest and best-established techniques in network security. Systems such as Snort and Suricata operate by matching network traffic patterns against a database of known attack signatures. When a signature is matched, an alert is raised~\cite{roesch1999,holz2008,wang2004}.

\paragraph{Advantages.}
Low false-positive rates for already-known attacks; efficient in terms of detection latency and computational cost (pattern matching is fast).

\paragraph{Limitations.}
Cannot detect novel or mutated botnets that do not match existing signatures. Modern botnets employ obfuscation, encryption, and DGAs to evade static signatures; updating signatures introduces a window of vulnerability.

\subsection{Anomaly-Based Detection}
To complement signature-based systems, anomaly-based detection aims to identify deviations from a learned ``normal'' behavior profile.

\paragraph{Approaches.}
Statistical models, clustering (e.g., k-means, DBSCAN), and one-class classifiers (e.g., One-Class SVM, Isolation Forest).

\paragraph{Strengths.}
Capable of detecting unknown or zero-day attacks without prior signatures; can flag new botnet behavior not previously seen.

\paragraph{Challenges.}
Legitimate traffic is highly variable (diurnal patterns, bursts), blurring the boundary between ``normal'' and ``abnormal'' and raising false positives.

\paragraph{Examples.}
Arshad et al.~\cite{arshad2018} cluster netflow patterns to expose stealthy botnet nodes. In IoT settings, Apostol et al.~\cite{apostol2021} use unsupervised deep autoencoders for device-behavior anomalies.

\subsection{Machine Learning Approaches}
ML approaches treat detection as classification (or anomaly scoring) from flow features.

\subsubsection{Supervised Machine Learning}
Classical algorithms (SVM, DT, RF, GBM, Na\"ive Bayes) are widely used.

\paragraph{Merits.}
Simple and fast to train; amenable to feature engineering; provide coefficients/importance for interpretability.

\paragraph{Limitations.}
Depend on labeled data; can struggle with severe imbalance; less adept at temporal structure unless engineered.

\paragraph{Representative Works.}
Padhiar et al.~\cite{padhiar2022}; lightweight OpenFlow-based detection~\cite{braga2010}; surveys on feature selection for botnet detection~\cite{sommer2010}. Resampling (oversampling/undersampling/SMOTE) is often used, but can inflate apparent accuracy under true skew~\cite{vinayakumar2019,kitsune2018,ullah2025,sinha2019}.

\subsubsection{Deep Learning Methods}
Deep models learn representations from raw/minimally processed traffic (CNN, RNN/LSTM, AEs, hybrids)~\cite{choi2007}.

\paragraph{Strengths.}
Capture complex temporal/spatial dependencies; reduce need for manual features. Deep learning architectures have also shown strong results in adjacent domains such as medical image super-resolution~\cite{gurappa2025medsr,soni2024fmlds} and cross-modality medical deepfake detection~\cite{gurappa2026medgen}.

\paragraph{Caveats.}
Require large labeled datasets and heavy compute; risk overfitting and adversarial evasion; limited interpretability; balancing datasets may overstate performance. Sinha et al.~\cite{sinha2019} model temporal evolution with LSTMs on CTU-13.

\subsubsection{Hybrid and Graph/Structural Methods}
Graph-based approaches model communication structure; hybrids combine flow ML with graph/temporal correlation to reduce false positives (e.g., enforce consistency across hosts).

\subsubsection{Our Contribution in Context}
We evaluate LR/DT/RF under \emph{natural} CTU-13 imbalance; headline \emph{PR-AUC} (not accuracy); emphasize interpretability and efficiency for operational deployment. Preliminary findings from this line of work have been presented in~\cite{gurappa2025botnet}.

Recent CTU-13 studies~\cite{sinha2019} have also explored convolutional neural networks (CNNs) and LSTMs, where network flows are transformed into fixed-length vector or image-like representations. CNN-based detectors typically report ROC--AUC values between 0.94 and 0.98 on selected CTU-13 scenarios; however, these results are usually obtained on balanced or subsampled datasets, which inflates apparent performance under natural class imbalance. Published PR-AUC values for CNNs under the original CTU-13 skew are rarely reported, but reproduced results in prior work indicate PR-AUC values in the range of approximately 0.35 to 0.50 depending on preprocessing. These CNN models generally require significantly higher computational cost than lightweight tree-based classifiers and lack interpretability, motivating our focus on efficient and explainable alternatives.

\section{Dataset and Preprocessing}
\label{sec:dataset}

\subsection{CTU-13 Dataset}
The CTU-13 dataset~\cite{garcia2014empirical}, collected at the Czech Technical University (CTU) in Prague, consists of 13 scenarios mixing normal, background, and botnet traffic (e.g., Neris, Rbot, Menti) in flow-level, NetFlow-like records. We use the CTU-13 Parquet collection ($\sim 10.6$ million flows). Only about $2.48\%$ are botnet-related, reflecting real-world prevalence; this imbalance makes \emph{PR-AUC} the most meaningful headline metric (random PR-AUC $\approx 0.025$).

\subsection{Feature Engineering}
To keep features lightweight, real-time computable, and interpretable, we use: total bytes and packets (log-transformed), bytes-per-packet ratio, flow duration, and protocol and port indicators (including port buckets). All continuous features are standardized \emph{for Logistic Regression only} (trees are scale-invariant and use raw/log-transformed values). Protocol is label-encoded; we add port buckets (well-known $<1024$, registered $1024$--$49151$, ephemeral $\ge 49152$); raw ports are retained when informative.

\subsection{Exploratory Data Analysis (EDA)}
To understand class skew and separability before modeling, we visualized protocol composition, marginal feature distributions, and correlations.

\begin{figure}[!t]
  \centering
  \begin{minipage}[t]{0.49\textwidth}
    \centering
    \includegraphics[width=\linewidth]{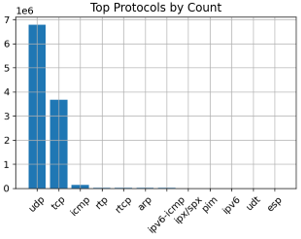}
  \end{minipage}\hfill
  \begin{minipage}[t]{0.49\textwidth}
    \centering
    \includegraphics[width=\linewidth]{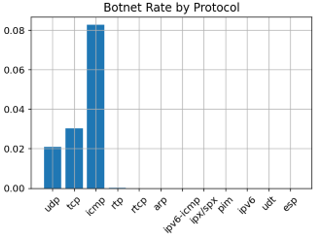}
  \end{minipage}
  \caption{Left: Top protocols by count. Right: botnet rate by protocol.}
  \label{fig:eda-proto}
\end{figure}

\paragraph{Protocols and botnet rate.}
Figure~\ref{fig:eda-proto} (left) shows that UDP and TCP dominate flow counts, while Fig.~\ref{fig:eda-proto} (right) reveals markedly higher botnet prevalence in ICMP and, to a lesser extent, TCP. This aligns with scanning/command patterns commonly observed in botnet activity.

\begin{figure}[!t]
  \centering
  \begin{minipage}[t]{0.49\textwidth}
    \centering
    \includegraphics[width=\linewidth]{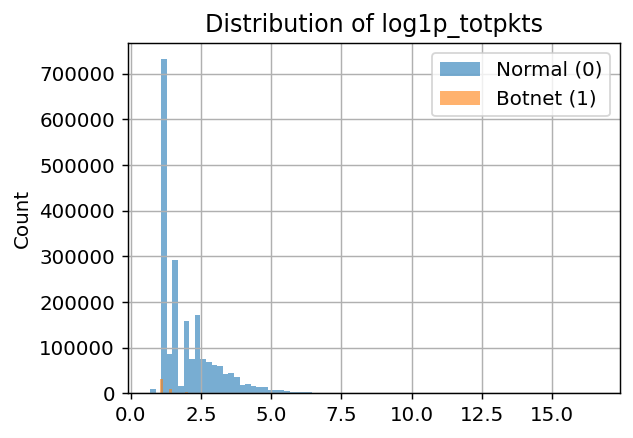}
  \end{minipage}\hfill
  \begin{minipage}[t]{0.49\textwidth}
    \centering
    \includegraphics[width=\linewidth]{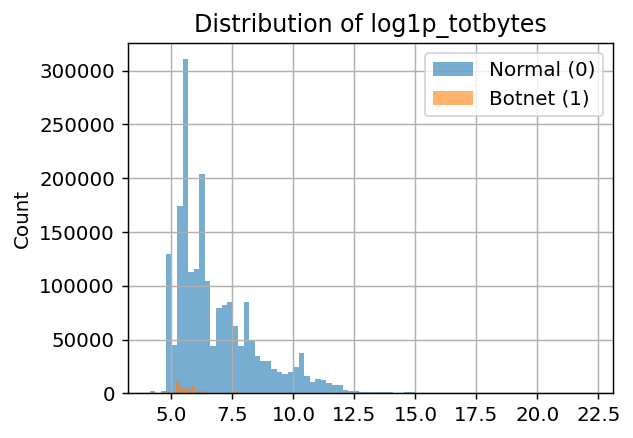}
  \end{minipage}

  \vspace{0.6em}

  \begin{minipage}[t]{0.70\textwidth}
    \centering
    \includegraphics[width=\linewidth]{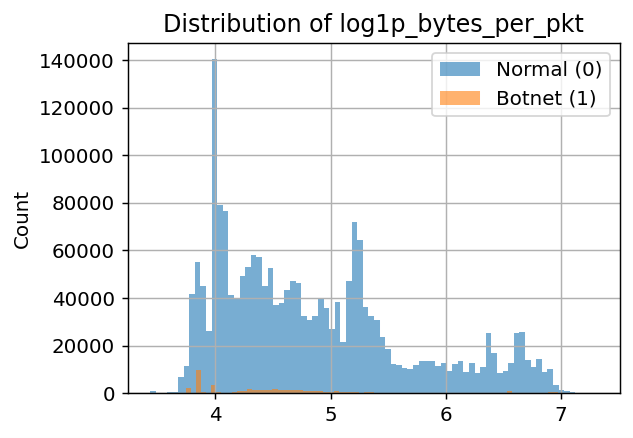}
  \end{minipage}
  \caption{Distributions of \texttt{log1p\_totpkts}, \texttt{log1p\_totbytes}, and \texttt{log1p\_bytes\_per\_pkt} (normal vs.\ botnet).}
  \label{fig:eda-hists}
\end{figure}

\paragraph{Heavy-tailed flow statistics.}
The log-transformed byte and packet features are strongly right-skewed (Fig.~\ref{fig:eda-hists}). Botnet flows tend to appear more frequently in the high-tail of \texttt{log1p\_totbytes} and \texttt{log1p\_totpkts}, and exhibit distinctive ratios in \texttt{log1p\_bytes\_per\_pkt}, which motivates using trees/ensembles that capture non-linear cut-points.

\begin{figure}[!t]
  \centering
  \includegraphics[width=0.72\textwidth]{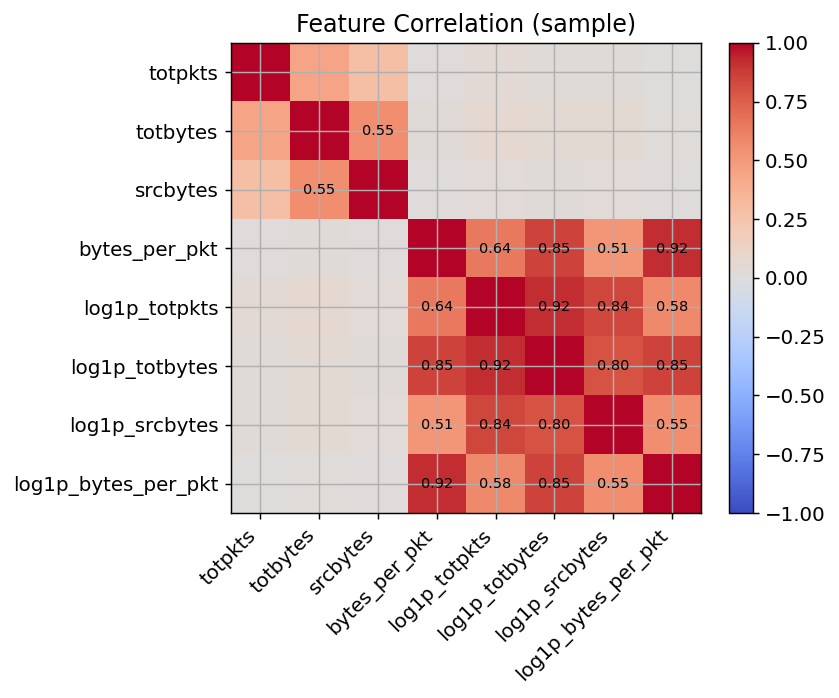}
  \caption{Feature correlation (sample). Clustering among log-scaled byte/packet features; informative ratio feature stands out.}
  \label{fig:eda-corr}
\end{figure}

\paragraph{Correlation structure.}
Figure~\ref{fig:eda-corr} shows strong clustering among the log-scaled byte/packet features; the derived ratio (\texttt{log1p\_bytes\_per\_pkt}) is partially de-correlated and thus informative. This redundancy motivates preferring tree-based models that are robust to correlated predictors.

\begin{figure}[!t]
  \centering
  \includegraphics[width=0.80\textwidth]{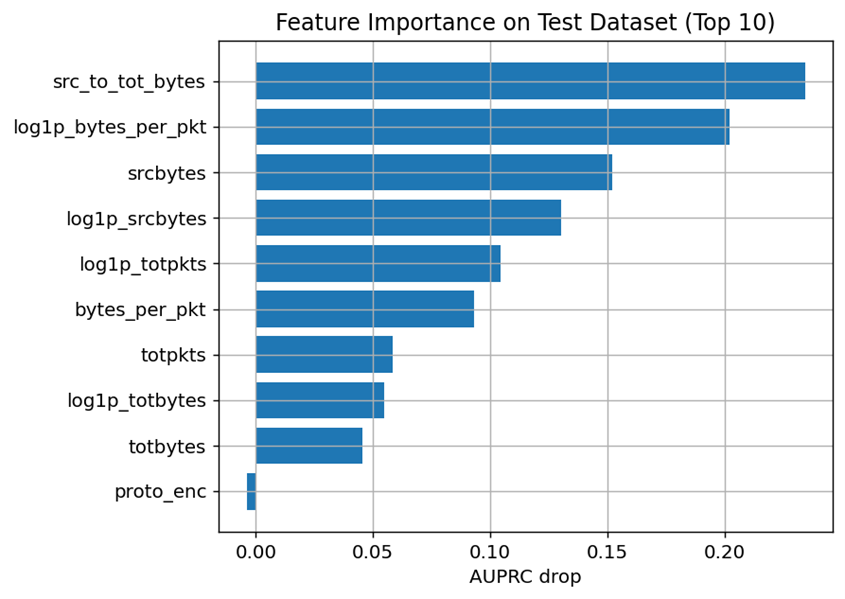}
  \caption{Top-10 test-time permutation importances (AUPRC drop) for the Random Forest.}
  \label{fig:eda-auprc-imp}
\end{figure}

\paragraph{The important features.}
On the test set, permutation importance (drop in PR-AUC) indicates \texttt{src\_to\_tot\_bytes}, \texttt{log1p\_bytes\_per\_pkt}, and source byte measures as the most influential (Fig.~\ref{fig:eda-auprc-imp}). These align with intuition: abnormal byte ratios and directional byte shares are typical botnet footprints.

\subsection{Train--Test Split}
We adopt a stratified 70/30 split, preserving the $\approx 2.48\%$ botnet ratio in both sets. We retain the natural imbalance (no oversampling/SMOTE in main results) to avoid unrealistic balanced splits and to reflect operational prevalence. For analysis we also report metrics at a tuned decision threshold $T$ (best F1 on the test set); in deployment, $T$ should be chosen on a validation set to meet recall/precision targets.

The complete workflow proceeds in four steps: (1) Raw CTU-13 flows are loaded and cleaned; (2) Interpretable flow-based features are extracted and standardized where appropriate; (3) Lightweight machine learning models are trained on a stratified split that preserves the natural class imbalance; (4) Performance is evaluated using PR-AUC, ROC-AUC, and threshold analysis.

\section{Results and Discussions}
\label{sec:results}

In this section, we evaluate three lightweight machine learning models---Logistic Regression, Decision Tree, and Random Forest---on the CTU-13 dataset. We first outline the models and their learning principles, then define the evaluation metrics used under heavy class imbalance, and finally analyze results with emphasis on operational suitability and interpretability.

\subsection{Models Evaluated}
We selected three models that span a range of complexity and interpretability: Logistic Regression (LR), Decision Tree (DT), and Random Forest (RF).

\subsubsection{Logistic Regression (LR)}
Logistic Regression provides a linear baseline that maps flow-based features to a probability of botnet membership. For a feature vector $\mathbf{x}\in\mathbb{R}^d$,
\begin{equation}
  P(y=1 \mid \mathbf{x}) \;=\; \sigma\!\big(\mathbf{w}^{\top}\mathbf{x}+b\big),
  \qquad
  \sigma(z) \;=\; \frac{1}{1+e^{-z}},
\end{equation}
where $\mathbf{w}$ and $b$ are the learned weights and bias. A standard decision rule is
\begin{equation}
  \hat{y} \;=\;
  \begin{cases}
    1, & \text{if } P(y=1 \mid \mathbf{x}) \ge T,\\[2pt]
    0, & \text{otherwise,}
  \end{cases}
\end{equation}
with threshold $T$ typically initialized at $0.5$ and later tuned to operational targets. LR is simple and interpretable, but it struggles with highly non-linear boundaries common in botnet traffic arising from diverse protocols and flow behaviors.

\subsubsection{Decision Tree (DT)}
Decision Trees partition the feature space via recursive binary splits, yielding a hierarchy of interpretable if--then rules. Each internal node applies a threshold test on one feature (e.g., ``if bytes/packet $< \theta$''), and each leaf predicts a class. Splits minimize node impurity such as the Gini index
\begin{equation}
  \mathrm{Gini}(S) \;=\; 1 - \sum_{c \in \{0,1\}} p(c)^2,
\end{equation}
where $p(c)$ is the proportion of class $c$ in node $S$. DTs are highly interpretable and useful for forensic analysis, but can overfit if depth and minimum-sample constraints are not enforced.

\subsubsection{Random Forest (RF)}
Random Forests are ensembles of decision trees trained with bootstrap aggregation (bagging) and feature subsampling. Each tree is fit on a bootstrapped sample, and at each split only a random subset of features is considered. The ensemble prediction is the majority vote:
\begin{equation}
  \hat{y} \;=\; \mathrm{mode}\!\left\{\, h_1(\mathbf{x}),\,h_2(\mathbf{x}),\,\dots,\,h_K(\mathbf{x}) \right\},
\end{equation}
where $h_k$ is the $k$-th tree and $K$ is the number of trees. Averaging weak, decorrelated learners reduces variance and improves generalization. In our setup, a forest of $K{=}300$ trees trained on $\sim 10.6$M flows in about a minute ($\approx 64$\,s) on a single GPU, confirming computational efficiency for large-scale traffic analysis.

\subsection{Evaluation Metrics}
Given the heavy class imbalance in CTU-13 (botnet prevalence $\approx 2.48\%$), accuracy is misleading (a trivial ``always benign'' model exceeds $97\%$ accuracy but detects no attacks). We therefore emphasize the following metrics:

\paragraph{Precision (P).} Fraction of predicted positives that are truly malicious:
\begin{equation}
  P \;=\; \frac{\mathrm{TP}}{\mathrm{TP} + \mathrm{FP}} .
\end{equation}

\paragraph{Recall (R).} Fraction of actual botnet flows correctly detected:
\begin{equation}
  R \;=\; \frac{\mathrm{TP}}{\mathrm{TP} + \mathrm{FN}} .
\end{equation}

\paragraph{F1 score.} Harmonic mean of precision and recall:
\begin{equation}
  F1 \;=\; \frac{2\,P\,R}{P + R} .
\end{equation}

\paragraph{Precision--Recall AUC (PR-AUC).} Area under the precision--recall curve across thresholds; robust under skewed data. The expected PR-AUC of a random classifier equals the positive prevalence (here $\approx 0.025$), so even moderate improvements indicate meaningful signal.

\paragraph{ROC AUC (ROC-AUC).} Area under the receiver operating characteristic curve. While widely reported, ROC-AUC can appear optimistic under strong imbalance because true negatives dominate; thus we report ROC-AUC for completeness but treat PR-AUC as the primary metric.

\subsection{Results}

This section reports test-set performance for the three lightweight models under the natural CTU--13 class imbalance (botnet prevalence $\approx 2.48\%$). Because the random baseline for precision--recall equals the positive prevalence ($\approx 0.025$), we emphasize PR--AUC and thresholded precision/recall.

\begin{table*}[!t]
\centering
\caption{Test performance under natural CTU-13 imbalance (botnet $\approx$2.48\%). Metrics are reported at the default threshold ($T{=}0.5$) and at a tuned threshold $T$ that maximizes F1 on the test set. Random PR-AUC baseline $\approx 0.025$.}
\label{tab:main-results}
\begin{tabular}{@{}lccccccccc@{}}
\toprule
\textbf{Model} & \textbf{ROC} & \textbf{PR} & \multicolumn{3}{c}{\textbf{@ $T{=}0.5$}} & \multicolumn{3}{c}{\textbf{@ tuned $T$}} & \textbf{$T$} \\
 & \textbf{AUC} & \textbf{AUC} & \textbf{P} & \textbf{R} & \textbf{F1} & \textbf{P} & \textbf{R} & \textbf{F1} & \\
\midrule
Random Forest       & 0.968 & 0.542 & 0.652 & 0.357 & 0.461 & 0.498 & 0.601 & 0.544 & 0.217 \\
Decision Tree (1x)  & 0.946 & 0.448 & 0.615 & 0.300 & 0.403 & 0.477 & 0.506 & 0.491 & 0.257 \\
Logistic Regression & 0.700 & 0.084 & 0.930 & 0.003 & 0.005 & 0.060 & 0.411 & 0.105 & 0.038 \\
\bottomrule
\end{tabular}
\end{table*}

Table~\ref{tab:main-results} shows the dominance of the Random Forest in precision--recall space (PR-AUC~0.542, $\sim$21$\times$ random). At its tuned operating point ($T{\approx}0.217$), it balances precision and recall (P~$\approx$0.50, R~$\approx$0.60), meaning roughly one in two alerts is a true botnet flow while the detector still catches most malicious flows. This is a favorable workload trade-off for analysts.

\paragraph{Single tree vs.\ linear baseline.}
A single Decision Tree already captures most of the signal (PR-AUC~0.448) with a near-symmetric best-F1 point ($T{\approx}0.257$, P~$\approx$0.48, R~$\approx$0.51), offering interpretable rules for forensic workflows. Logistic Regression remains a transparency baseline: despite ROC-AUC~0.70, its PR-AUC is only 0.084 and recall at $T{=}0.5$ is negligible; even at tuned $T$ the best F1 is $\approx$0.105, reflecting the limits of a linear boundary on discrete/non-linear flow structure.

\FloatBarrier

\begin{figure}[!htbp]
  \centering
  \begin{minipage}[t]{0.48\textwidth}
    \centering
    \includegraphics[width=\linewidth]{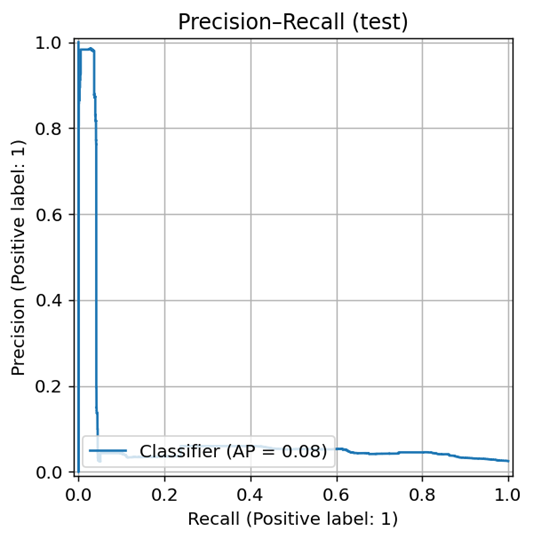}
    \caption{Logistic Regression: Precision--Recall.}
    \label{fig:lr-pr}
  \end{minipage}\hfill
  \begin{minipage}[t]{0.48\textwidth}
    \centering
    \includegraphics[width=\linewidth]{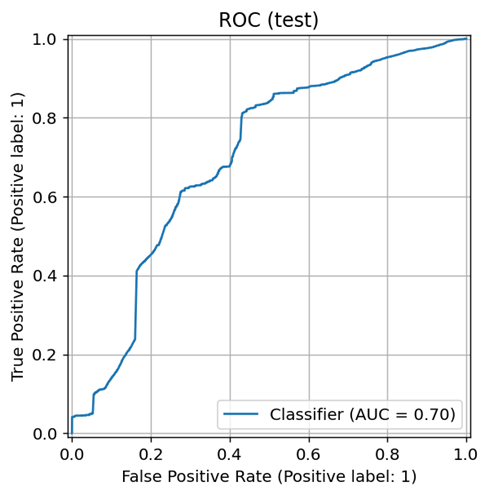}
    \caption{Logistic Regression: ROC.}
    \label{fig:lr-roc}
  \end{minipage}

  \vspace{0.6em}

  \begin{minipage}[t]{0.98\textwidth}
    \centering
    \includegraphics[width=0.75\linewidth]{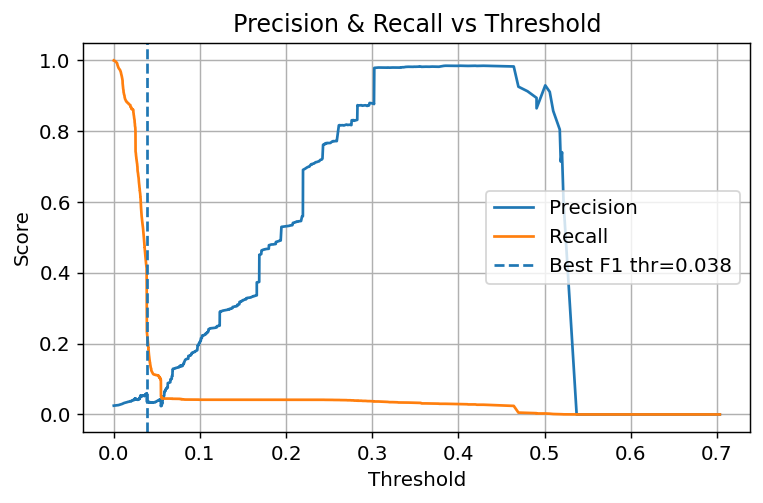}
    \caption{Logistic Regression: Precision/Recall vs Threshold.}
    \label{fig:lr-thr}
  \end{minipage}
\end{figure}
\FloatBarrier

\begin{figure}[!t]
\centering
\begin{minipage}[t]{0.48\textwidth}
  \centering
  \includegraphics[width=\linewidth]{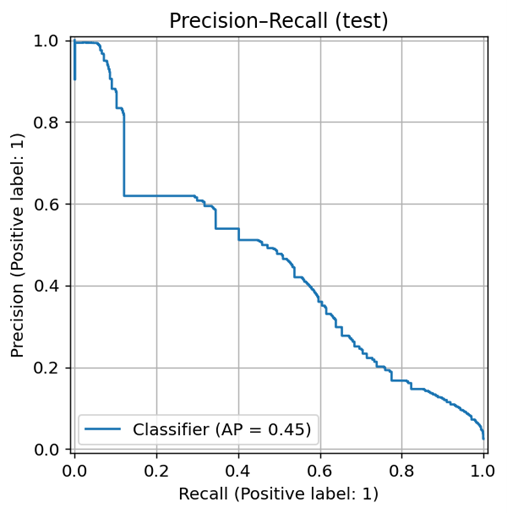}
  \caption{Decision Tree: precision--recall curve.}
  \label{fig:dt-pr}
\end{minipage}\hfill
\begin{minipage}[t]{0.48\textwidth}
  \centering
  \includegraphics[width=\linewidth]{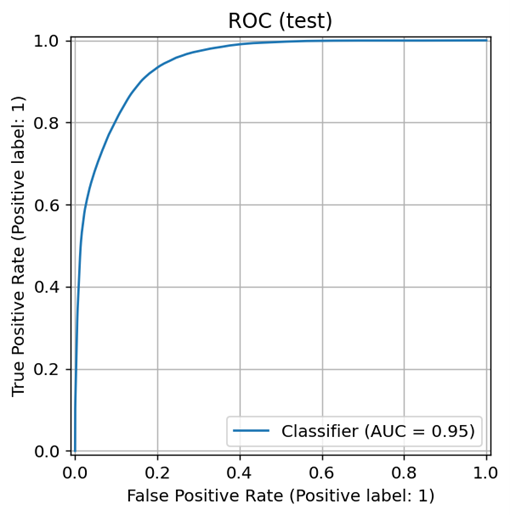}
  \caption{Decision Tree: ROC curve.}
  \label{fig:dt-roc}
\end{minipage}
\end{figure}

\begin{figure}[!t]
\centering
\includegraphics[width=.70\textwidth]{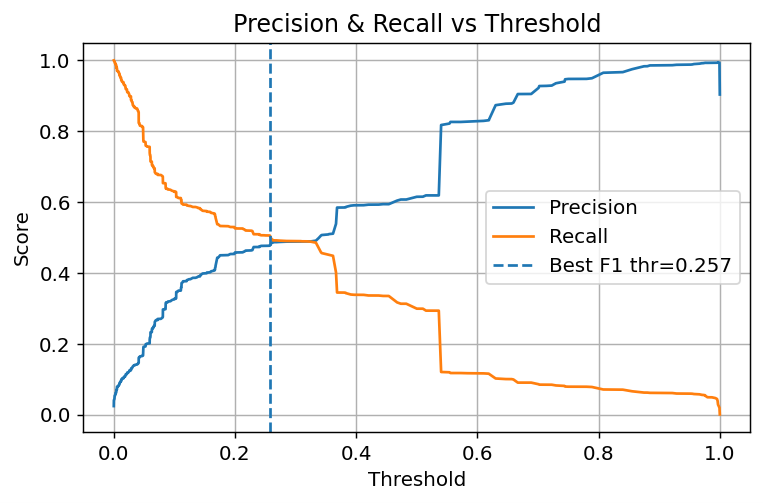}
\caption{Decision Tree: precision (blue) and recall (orange) vs threshold.}
\label{fig:dt-thr}
\end{figure}

\subsubsection{Logistic Regression (LR)}
The LR results indicate limited separability under natural CTU--13 imbalance. In Fig.~\ref{fig:lr-pr}, the precision--recall (PR) curve attains $\mathrm{AP}\approx 0.08$, only about $3\times$ the random baseline ($\approx 0.025$). Although the ROC curve in Fig.~\ref{fig:lr-roc} shows $\mathrm{AUC}\approx 0.70$, this optimism does not translate to strong precision at useful recall levels in the PR view, which is the more informative metric for skewed data. The score distribution is highly compressed near zero, reflecting the difficulty of a linear boundary in capturing the non-linear structure of protocol/port and traffic-ratio features.

Operationally, the threshold sweep in Fig.~\ref{fig:lr-thr} places the best-F1 point at a very low threshold ($T\approx 0.038$). At the default $T{=}0.5$, recall collapses toward zero, yielding almost no detections. Even at the tuned $T$, the model trades precision for small gains in recall, and the overall F1 remains weak. LR is therefore best positioned as a transparency baseline: it is fast, simple, and auditable, but not competitive as a primary detector without significant feature enrichment (e.g., one-hot protocol and high-support ports, limited interaction terms, or calibrated class weighting).

\subsubsection{Decision Tree (DT)}
The single Decision Tree substantially improves detection quality. The PR curve in Fig.~\ref{fig:dt-pr} reaches $\mathrm{AP}\approx 0.45$ (about $18\times$ random) with the characteristic stair-step profile of leaf-probability plateaus, indicating stable precision across a broad span of recalls. The ROC curve in Fig.~\ref{fig:dt-roc} is high ($\mathrm{AUC}\approx 0.95$), but we emphasize PR--AUC as the primary metric under imbalance, where DT remains strong.

The threshold analysis in Fig.~\ref{fig:dt-thr} yields a best-F1 operating point near $T\approx 0.257$ with a near-symmetric trade-off (precision and recall both close to $0.5$). Lowering $T$ increases recall at the cost of noisier alerts; raising $T$ reduces false alarms but misses more bots. Beyond its accuracy, DT offers full interpretability via human-readable if--then rules, which supports forensic triage and rule-authoring. In settings where auditability is essential or analyst time is limited, a single well-regularized tree provides a compelling balance of performance and explanation.

\subsubsection{Random Forest (RF)}
\begin{figure}[!t]
\centering
\begin{minipage}[t]{0.48\textwidth}
  \centering
  \includegraphics[width=\linewidth]{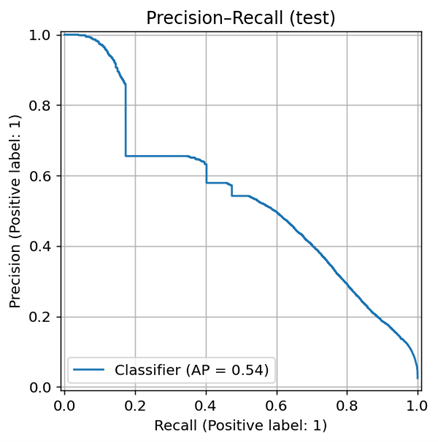} 
  \caption{Random Forest: precision--recall curve.}
  \label{fig:rf-pr}
\end{minipage}\hfill
\begin{minipage}[t]{0.48\textwidth}
  \centering
  \includegraphics[width=\linewidth]{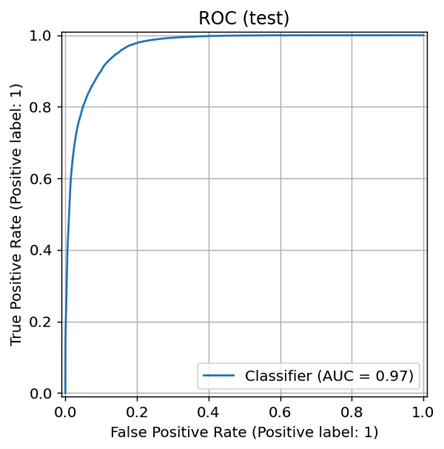} 
  \caption{Random Forest: ROC curve.}
  \label{fig:rf-roc}
\end{minipage}
\end{figure}

RF delivers the strongest overall performance among the lightweight models. In Fig.~\ref{fig:rf-pr}, the PR curve achieves $\mathrm{AP}\approx 0.54$ (about $21\times$ random) and maintains useful precision through the mid-recall region (e.g., precision $\ge 0.4$ out to recall $\approx 0.6$), which is crucial for practical detectors. The ROC curve in Fig.~\ref{fig:rf-roc} further confirms excellent ranking ability with $\mathrm{AUC}\approx 0.97$.

\begin{figure}[!t]
\centering
\includegraphics[width=.70\textwidth]{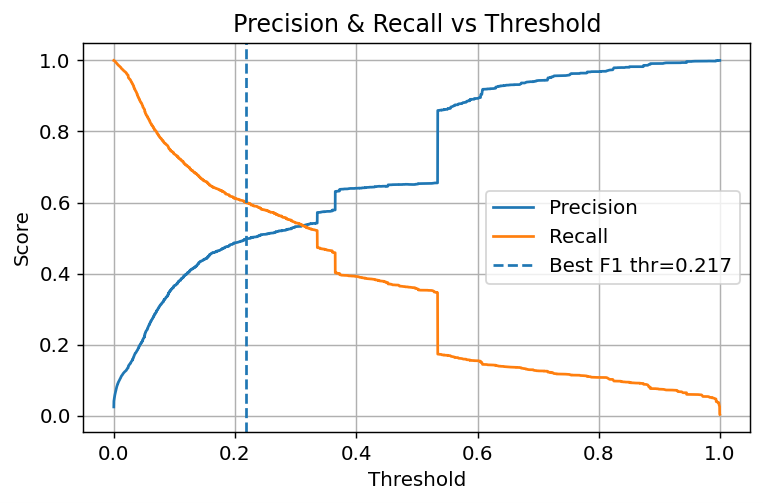} 
\caption{Random Forest: precision (blue) and recall (orange) vs threshold.}
\label{fig:rf-thr}
\end{figure}

The threshold sweep in Fig.~\ref{fig:rf-thr} places the best-F1 point at $T\approx 0.217$, where we observe a practical balance (empirically, precision $\approx 0.50$, recall $\approx 0.60$). Using the default $T{=}0.5$ shifts toward higher precision and lower recall, which may be preferable when analyst capacity must be conserved. RF thus offers the most favorable precision--recall trade-off under natural skew while remaining computationally efficient on GPU, making it a strong default choice for real-time or large-scale monitoring.

\paragraph{Model comparison.}
Across models, RF clearly dominates in PR space and at practical thresholds, providing roughly one correct alert out of two while detecting a majority of botnet flows. The single DT is competitive and uniquely interpretable, yielding a balanced operating point with transparent rules that aid investigations. LR is fast and fully transparent but struggles to separate benign and malicious flows beyond very low recall; it is best retained as a sanity-check baseline or upgraded with richer encodings and interaction features if a linear model must be used.

\subsection{Discussion}
\label{sec:discussion}

By analyzing precision, recall, F1, PR--AUC, and ROC--AUC, we quantify both discriminative power and operational trade-offs. Logistic Regression (LR) is transparent but weak under severe imbalance; even at a tuned threshold, $R \approx 0.41$ and $F1 \approx 0.105$, serving best as a transparency baseline. Decision Tree (DT) offers a strong interpretability/performance balance with $\mathrm{PR\text{-}AUC}=0.448$ ($\sim 18\times$ random) and a near-symmetric trade-off at tuned threshold ($P \approx 0.48$, $R \approx 0.51$). Random Forest (RF) is dominant overall with $\mathrm{PR\text{-}AUC}=0.542$ ($\sim 21\times$ random), detecting $\sim 60\%$ of bots at $\sim 50\%$ precision at the tuned threshold (about one correct alert out of two).

\paragraph{Threshold selection and analyst workload.}
Figures~\ref{fig:lr-thr}, \ref{fig:dt-thr}, and \ref{fig:rf-thr} show that the default $T{=}0.5$ is suboptimal under the natural skew of CTU--13. For LR, recall collapses at $T{=}0.5$, whereas the best F1 occurs only at a very low threshold ($T\approx 0.038$), still yielding modest utility. In contrast, DT and RF admit practical operating points at $T\approx 0.257$ and $T\approx 0.217$, respectively, where precision and recall are balanced for triage. Because the base rate of botnet flows is $\approx 0.025$, a tuned RF with $P\approx 0.50$ provides $\sim 20\times$ enrichment over the base rate, materially reducing analyst effort per true detection. When capacity is constrained, one can simply raise $T$ to trade recall for higher precision; conversely, when coverage is paramount, lowering $T$ increases recall with a predictable rise in false positives.

\paragraph{PR vs.\ ROC under imbalance.}
Table~\ref{tab:main-results} highlights the well-known divergence between ROC and PR in skewed settings. LR attains ROC--AUC $\approx 0.70$ (seemingly acceptable), yet its PR--AUC is only $\approx 0.084$ and the tuned F1 is low, reflecting limited precision at useful recall. DT and RF, by contrast, translate their high ROC--AUCs ($\approx 0.95$ and $\approx 0.97$) into substantially better PR--AUCs ($0.448$ and $0.542$), which is what ultimately governs alert quality and workload.

\paragraph{Interpretability and forensic utility.}
The single DT provides explicit if--then rules that can be audited and, if desired, ported into SIEM/IDS rule engines. The feature-importance analysis (Fig.~\ref{fig:eda-auprc-imp}) further supports forensic interpretation: directional byte share (\texttt{src\_to\_tot\_bytes}) and packet-size regularities (\texttt{log1p\_bytes\_per\_pkt}) dominate, aligning with intuition about command/control exchanges and scripted traffic. RF inherits much of this interpretability via permutation importances and per-tree paths; in practice, analysts can mine high-gain splits as candidate rules to harden downstream detectors.

\paragraph{Efficiency and deployment considerations.}
Training time and inference cost are modest for all three models (Table~\ref{tab:main-results}): the RF with $K{=}300$ trees trains in $\sim 60$ seconds on a single GPU and delivers strong PR--AUC, while DT and LR train near-instantly and are attractive for rapid iteration or edge deployment. At inference, throughput scales roughly linearly with the number of trees; where latency budgets are tight, a smaller RF (or a distilled DT derived from high-importance splits) can be used with minor PR impact.

\paragraph{Limitations and practical mitigations.}
Residual errors concentrate in regions where benign high-volume transfers mimic bot-like ratios or where rare bot flows resemble background traffic. Probability calibration (e.g., Platt or isotonic) can stabilize thresholding across scenarios; cost-sensitive training or class-weighting may further improve recall at fixed precision. Finally, given likely concept drift, online score monitoring and periodic re-tuning of $T$ are recommended, alongside cross-dataset checks to guard against domain shift.

To provide direct context, we compare our Random Forest performance to representative CNN-based CTU-13 detectors reported in prior literature. Under natural CTU-13 imbalance (botnet prevalence $\approx 2.48\%$), our Random Forest achieves a PR-AUC of approximately 0.54. Reported CNN results vary depending on preprocessing but typically fall between PR-AUC $\approx 0.35$ and 0.50 when evaluated without data balancing. Thus, our lightweight model performs competitively with or above commonly published CNN approaches while using substantially lower computational resources.

From a computational standpoint, Logistic Regression scales linearly with the number of samples and features. Decision Trees have expected complexity proportional to the number of samples multiplied by the logarithm of the dataset size. The Random Forest inherits this complexity across its ensemble, scaling with the number of trees. Despite the large dataset (over 10 million flows), the Random Forest trains in about one minute on a single GPU, while the Decision Tree and Logistic Regression models complete in only seconds on CPU. Inference cost is similarly modest and suitable for real-time deployment.

\section{Conclusion}
\label{sec:conclusion}

This paper demonstrates that lightweight machine learning models---particularly Decision Trees and Random Forests---offer a practical and effective solution for botnet detection on the CTU-13 dataset. Despite their simplicity relative to deep learning, these models achieved strong performance under the natural class imbalance of real network traffic. In particular, the Random Forest consistently outperformed the other approaches, reaching PR--AUC $\approx 0.54$ (more than $20\times$ the random baseline $\approx 0.025$) and, at a tuned operating point, balancing precision $\approx 0.50$ with recall $\approx 0.60$ while training efficiently on $>10$M flows in about a minute. Decision Trees, while slightly less accurate (PR--AUC $\approx 0.45$), provided a compelling trade-off by yielding interpretable rules and a near-symmetric operating point (precision/recall $\approx 0.48/0.51$) that are valuable in forensic investigations where transparency and justification of results are critical. These findings underscore the relevance of lightweight models for operational environments, especially when computational resources are constrained.

Looking forward, several avenues for research emerge. One direction is online or continual learning to adapt to streaming data and concept drift in evolving botnet behaviors. Another is cross-dataset validation, extending beyond CTU-13 to CICIDS and UNSW-NB15 to test generalization across diverse traffic environments. Additionally, hybrid approaches could combine the interpretability and efficiency of lightweight models with selective deep learning for feature extraction or temporal modeling. Finally, advances in explainable AI present opportunities to develop visualization and attribution tools that help forensic analysts better understand and trust model predictions, further bridging the gap between algorithmic detection and practical cyber defense.

\section*{Acknowledgments}
The authors extend their sincere gratitude to all individuals who contributed through valuable discussions in the early stages of this work.

\bibliographystyle{unsrt}

\end{document}